\documentclass[12pt]{article}%
\usepackage{amsmath}
\usepackage{graphicx}
\usepackage{amsfonts}
\usepackage{amssymb}%
\providecommand{\U}[1]{\protect\rule{.1in}{.1in}}
\begin{document}

\begin{center}
{\Large \textbf{Modified Hamiltonian for a particle in an infinite box
\\[0pt]}}

{\Large \textbf{that includes wall effects\\[0pt]}}
\end{center}

\baselineskip13.5pt

\begin{center}
{\large \textbf{\\[0pt]}}

{\ Leon Cohen}$^{a}$, {\ Rafael Sala Mayato}$^{b}$, {\ Patrick Laughlin}$^{c}$

$^{a}$\textit{Department of Physics,\\[0pt]Hunter College of the City
University of New York, 695 Park Ave., New York, NY 10065, USA\\[0pt]}

$^{b}$\textit{Departamento de F\'\i sica and IUDEA,\\[0pt]Universidad de La
Laguna, La Laguna, 38204, S/C de Tenerife, Spain\\[0pt]}

$^{c}$\textit{Departments of Bioengineering, and Electrical \& Computer
Engineering,\\[0pt]University of Pittsburgh, Pittsburgh, PA15261, USA\\[0pt]}

$\ $ \\[0pt]
\end{center}

\hskip                    .40cm

\vskip.14in \hskip.5cm\begin{minipage}[t]{15.05cm}
\noindent{We give a modified Hamiltonian for a particle in a box with infinite potential
walls that takes into account wall effects. The Hamiltonian is expressed in both the position and momentum representation.
In the momentum representation the eigenvalue problem for energy is an integral equation.
}
\end{minipage}\vskip.2truein \noindent Keywords: {\small \ particle in a box;
modified Hamiltonian}

\vskip.3in

\begin{center}
{\large \textbf{1. Introduction}}
\end{center}

The usual textbook approach for solving the eigenvalue problem for energy for
a particle in a box with infinite potential at the walls
\begin{equation}
\mathbf{H}u(x)=Eu(x)
\end{equation}
is to write the Hamiltonian,%
\begin{equation}
\mathbf{H}=\frac{\mathbf{p}^{2}}{2m}\label{eq2}%
\end{equation}
and in the position representation, the eigenvalue problem is taken to be
\begin{equation}
-\frac{\hbar^{2}}{2m}\frac{d^{2}}{dx^{2}}u(x)=Eu(x)
\end{equation}
and the \textquotedblleft solution\textquotedblright\ \ is then written as%
\begin{equation}
u(x)=A\sin\sqrt{\frac{2mE}{\hbar^{2}}}x+B\cos\sqrt{\frac{2mE}{\hbar^{2}}}x
\end{equation}
Upon imposing the usual boundary conditions
\begin{equation}
u(0)=u(L)=0
\end{equation}
one obtains the normalized eigenfunctions
\begin{equation}
u_{n}(x)=\sqrt{\frac{2}{L}}\sin\left(  \frac{n\pi}{L}x\right)  \hspace
{0.2in}0\leq x\leq L
\end{equation}
with eigenvalues%
\begin{equation}
E_{n}=\frac{\pi^{2}\hbar^{2}}{2mL^{2}}n^{2}\hspace{0.2in}n=1,2,....\label{e}%
\end{equation}
The $u_{n}(x)$'s are normalized and orthonormal%

\begin{equation}
\int_{0}^{L}u_{n}(x)u_{m}(x) dx=\delta_{nm}\;\;\;\;\;
\end{equation}

Writing the solutions and Hamiltonian and hence the eigenfunctions in the
above forms leads to many difficulties. There is a vast literature on the
subject and here we just touch on some of the issues [1-12]. If the
Hamiltonian is given by Eq. (\ref{eq2}), it would seem that $\mathbf{H}$ and
the momentum operator, $\mathbf{p,}$ commute and would hence have the same
eigenfunctions, which is clearly not the case. Also, for any particular state
of energy we have that $\left\langle \mathbf{H}^{2}\right\rangle =\left\langle
\mathbf{H}\right\rangle ^{2}$ which of course is correct. However, if one
substitutes Eq. (\ref{eq2})\ into this one obtains that $\left\langle
\mathbf{p}^{4}\right\rangle $ is equal to $\left\langle \mathbf{p}%
^{2}\right\rangle ^{2}$ which is incorrect since $\left\langle \mathbf{p}%
^{2}\right\rangle $ is finite but $\left\langle \mathbf{p}^{4}\right\rangle
=\infty$ \cite{roja}. Additionally, some have argued that $\left\langle
p\right\rangle =\pm\sqrt{2mE_{n}}$ and sometimes this is interpreted as that
the particle moves to the right or left with one of these values. This is
false, since all values of momentum are possible.\cite{bell,mark,rigg}

These difficulties and others are resolved by a modified Hamiltonian that
takes into account the wall effects.

\begin{center}
\bigskip

{\large \textbf{2. \ Modified Hamiltonian}}
\end{center}

We consider the following modified Hamiltonian, $\mathbf{H}_{M},$ in the
position representation
\begin{equation}
\mathbf{H}_{M}=\frac{\hbar^{2}}{2m}\frac{d^{2}}{dx^{2}}+\frac{\hbar^{2}}%
{2m}\frac{\left[  \delta{(x)-\delta(x-L)}\right]  }{\left[  {\Theta
(x)-\Theta(x-L)}\right]  }\frac{d}{dx}\label{hnew}%
\end{equation}
where
\begin{equation}
{\Theta(x)}=%
\begin{cases}
0 & \text{if }x<0,\\
1 & \text{if }x\geq0.
\end{cases}
\end{equation}

The eigenvalue problem for $\mathbf{H}_{M}$, is
\begin{equation}
\mathbf{H}_{M}v_{n}(x)=E_{n}v_{n}(x)
\end{equation}
where $v_{n}(x)$ are the eigenfunctions. Explicitly
\begin{equation}
-\frac{\hbar^{2}}{2m}\frac{d^{2}}{dx^{2}}v_{n}+\frac{\hbar^{2}}{2m}%
\frac{\left[  \delta{(x)-\delta(x-L)}\right]  }{\left[  {\Theta(x)-\Theta
(x-L)}\right]  }\frac{d}{dx}v_{n}=E_{n}v_{n}\label{hnew2}%
\end{equation}
The solutions, that is the eigenfunctions, are
\begin{align}
v_{n}(x)  &  =\left[  {\Theta(x)-\Theta(x-L)}\right]  \sqrt{\frac{2}{L}}%
\sin\left(  \frac{n\pi}{L}x\right) \label{vnew}\\
&  =\left[  {\Theta(x)-\Theta(x-L)}\right]  u_{n}(x)
\end{align}
with $E_{n}$ given by Eq. (\ref{e}).

Also, we point out that the operator defined by
\begin{equation}
\mathbf{H}_{M^{\prime}}=\frac{\hbar^{2}}{2m}\frac{d^{2}}{dx^{2}}+\frac
{\hbar^{2}}{2m}\left[  \delta{(x)-\delta(x-L)}\right]  \frac{d}{dx}%
\end{equation}
when operating on an eigenfunction, $v_{n}(x),$ is equivlent to $\mathbf{H}%
_{M}$ operating on $v_{n}(x)$, that is $\mathbf{H}_{M}v_{n}(x)=\mathbf{H}%
_{M^{\prime}}v_{n}(x)$

In the momentum representation the eigenvalue problem becomes an integral
equation
\begin{equation}
\frac{p^{2}}{2m}\varphi_{n}(p)+\frac{i}{2m}\frac{1}{2\pi}\int_{-\infty
}^{\infty}(1-e^{-iL(p-p^{\prime})/\hbar})p^{\prime}\varphi_{n}(p^{\prime
})dp^{\prime}=E_{n}\varphi_{n}(p)\label{mom}%
\end{equation}
where%
\begin{align}
\varphi_{n}(p) &  =\frac{1}{\sqrt{2\pi\hbar}}\int_{-\infty}^{\infty}%
v_{n}(x)e^{-ixp/\hbar}dx\label{m1}\\
v_{n}(x) &  =\frac{1}{\sqrt{2\pi\hbar}}\int_{-\infty}^{\infty}\varphi
_{n}(p)e^{ixp/\hbar}dp\label{m2}%
\end{align}
The solution to Eq. (\ref{mom}), are found to be
\begin{equation}
\varphi_{n}(p)=\frac{\pi}{L\sqrt{\pi\hbar L}}\frac{n}{\left(  \frac{n\pi}%
{L}\right)  ^{2}-\left(  \frac{p}{\hbar}\right)  ^{2}}\left[  1-(-1)^{n}%
e^{-ipL/\hbar}\right]
\end{equation}
which may be obtained directly from Eq. (\ref{m1})\ \cite{bell,mark}. An
alternative expression for $\varphi_{n}(p)$ is
\begin{equation}
\varphi_{n}(p)=\frac{1}{2i}\sqrt{\frac{L}{\pi\hbar}}\left\{  e^{-iL(\frac
{p}{\hbar}-\frac{n\pi}{L})/2}\frac{\sin\left(  L(\frac{p}{\hbar}-\frac{n\pi
}{L})/2\right)  }{L(\frac{p}{\hbar}-\frac{n\pi}{L})/2}-e^{-iL(\frac{p}{\hbar
}+\frac{n\pi}{L})/2}\frac{\sin\left(  L(\frac{p}{\hbar}+\frac{n\pi}%
{L})/2\right)  }{L(\frac{p}{\hbar}+\frac{n\pi}{L})/2}\right\}
\end{equation}
The probability distribution for momentum, $\left\vert \varphi_{n}%
(p)\right\vert ^{2},$ is given by%

\begin{equation}
\left\vert \varphi_{n}(p)\right\vert ^{2}=\frac{4n^{2}\pi L/\hbar}{\left(
\left(  n\pi\right)  ^{2}-\left(  pL/\hbar\right)  ^{2}\right)  ^{2}}\left\{
\begin{array}
[c]{cc}%
\sin^{2}\frac{pL}{2\hbar} & n=\text{even}\\
\cos^{2}\frac{pL}{2\hbar} & n=\text{odd}%
\end{array}
\right.
\end{equation}
or
\begin{equation}
\left\vert \varphi_{n}(p)\right\vert ^{2}=\frac{8m\hbar E_{n}}{\pi L}\frac
{1}{\left(  p^{2}-2mE_{n}\right)  ^{2}}\left\{
\begin{array}
[c]{cc}%
\sin^{2}\frac{pL}{2\hbar} & n=\text{even}\\
\cos^{2}\frac{pL}{2\hbar} & n=\text{odd}%
\end{array}
\right.
\end{equation}

\begin{center}
\bigskip

{\large \textbf{3. Discussion and Conclusion}}

\smallskip
\end{center}

The results presented in the last section resolve the difficulties discussed
in the introduction and they also resolve other issues. This will be discussed
in an expanded version of this paper. We point out here that since the
Hamiltonian is no longer given by Eq. (\ref{eq2})\ we can no longer conclude
that $\left\langle \mathbf{p}^{4}\right\rangle $ is equal to $\left\langle
\mathbf{p}^{2}\right\rangle ^{2}$ $.$ Calculation of $\left\langle
\mathbf{p}^{2}\right\rangle $ using Eq. (\ref{vnew})\ does give the correct
answer while calculation of $\left\langle \mathbf{p}^{4}\right\rangle $ with
Eq. (\ref{vnew}) does indeed give infinity. Note, too, that since the modified
Hamiltonian $\mathbf{H}_{M}$ and $\mathbf{p}$ do not commute one can not use
$\left\langle \mathbf{H}_{M}^{k}\right\rangle $ to calculate $\left\langle
\mathbf{p}^{2k}\right\rangle ,$ but direct calculation of $\left\langle
\mathbf{p}^{2k}\right\rangle $ with the new eigenfunctions, $v_{n},$ does give
the correct answers, namely%
\begin{equation}
\left\langle \mathbf{p}^{2k}\right\rangle =\int v_{n}(x)\mathbf{p}^{2k}%
v_{n}(x)dx=%
\begin{cases}
2mE_{n} & \text{for }k=1,\\
\infty & \text{for }k\geq2.
\end{cases}
\end{equation}

\end{document}